\documentclass[conference]{IEEEtran}
\IEEEoverridecommandlockouts
\usepackage{cite}
\usepackage{amsmath,amssymb,amsfonts}
\usepackage{algorithmic}
\usepackage{graphicx}
\usepackage{textcomp}
\usepackage{xcolor}
\usepackage{url}
\usepackage{rotating}
\usepackage{multirow}

\usepackage[nameinlink,noabbrev]{cleveref} 
\crefname{section}{Section}{Sections}
\crefname{subsection}{Subsection}{Subsections}
\crefname{figure}{Figure}{Figures}
\crefname{table}{Table}{Tables}

\def\BibTeX{{\rm B\kern-.05em{\sc i\kern-.025em b}\kern-.08em

    T\kern-.1667em\lower.7ex\hbox{E}\kern-.125emX}}
\begin{document}

\title{End-to-end NILM System Using High Frequency Data and Neural Networks\\
}

\author{

\IEEEauthorblockN{Marchesoni-Acland\IEEEauthorrefmark{1}}
\IEEEauthorblockA{\textit{Instituto de Ingeniería Eléctrica} \\
\textit{Universidad de la República}\\
Montevideo, Uruguay \\
marchesoni@fing.edu.uy}
\and
\IEEEauthorblockN{Mari\~{n}o\IEEEauthorrefmark{1}}
\IEEEauthorblockA{\textit{Instituto de Ingeniería Eléctrica} \\
\textit{Universidad de la República}\\
Montevideo, Uruguay \\
cmarino@fing.edu.uy}
\and

\IEEEauthorblockN{Masquil\IEEEauthorrefmark{1}}
\IEEEauthorblockA{\textit{Instituto de Ingeniería Eléctrica} \\
\textit{Universidad de la República}\\
Montevideo, Uruguay \\
eliasmasquil@gmail.com}  \linebreakand 
\IEEEauthorblockN{Masaferro}
\IEEEauthorblockA{\textit{Instituto de Ingeniería Eléctrica} \\
\textit{Universidad de la República}\\
Montevideo, Uruguay \\
pmasaferro@fing.edu.uy}
\and

\IEEEauthorblockN{Fernández}
\IEEEauthorblockA{\textit{Instituto de Ingeniería Eléctrica} \\
\textit{Universidad de la República}\\
Montevideo, Uruguay \\
alicia@fing.edu.uy}

\thanks{\IEEEauthorrefmark{1}These authors equally contributed.}

}


\maketitle

\begin{abstract}
Improving energy efficiency is a necessity in the fight against climate change.
Non Intrusive Load Monitoring (NILM) systems give important information about the household consumption that can be used by the electric utility or the end users.
In this work the implementation of an end-to-end NILM system is presented, which comprises a custom high frequency meter and neural-network based algorithms.
The present article presents a novel way to include high frequency information as input of neural network models by means of multivariate time series that include carefully selected features.
Furthermore, it provides a detailed assessment of the generalization error and shows that this class of models generalize well to new instances of seen-in-training appliances.
An evaluation database formed of measurements in two Uruguayan homes is collected and discussion on general unsupervised approaches is provided.
\end{abstract}


\section{Introduction}
Climate change is a consequence of greenhouse gas emissions and causes more extreme climate conditions that imply severe negative effects.
According to the Intergovernmental Panel on Climate Change (IPCC), electricity and heat production accounts for a quarter of total global emissions.
To limit temperature growth to $1.5$ K ambitious efforts have to be carried out, including improving energy efficiency.
Energy efficiency implies reducing the consumption for a given comfort or production level.
Information availability is crucial to improve in this line.
Non intrusive load monitoring (NILM) systems were introduced by Hart \cite{hart1992nonintrusive} and their objective is to get valuable information of consumption in an electric installation from measurements taken at only one point.

This work presents an end-to-end NILM system. This implies building or getting a meter, and using some disaggregation algorithm. The meter is usually connected to the main electrical panel of a house. The meter's goal is to measure, directly or indirectly, the power consumption of the house. A common solution is to use the electric utility's smart meters, with the drawback of a low sample rate, $1.1$ mHz in Uruguay.
Another direct solution is to buy some comercial meter. The most common drawbacks are the low sample rate and lack of flexibility in its use.
These meters are often equipped with a communication system that allows data transmission to other points, via Ethernet, WiFi, or IoT oriented protocols as MQTT or ZigBee.
The maximum sample rate of comercial meters is about $1$ Hz \cite{rafsanjani2018linking, jin2017virtual}. The other options are designing a custom meter, as was done in this work, or purchasing a commercial acquisition board. For more details see \cite{ruano2019nilm}.

The other axis of the solution refers to dissagregation methods. Classical algorithms are Combinatorial Optimization and Factorial Hidden Markov Chains \cite{kelly2014NILMTKv02}, and an important reference for this work that introduces Artificial Neural Networks (ANNs) for the problem is \cite{Kelly2015}. In that work ANNs show promise, as it is reported that their generalization capabilities are good. In fact, performance seems to be better for unknown appliances than for appliances seen in training. This work replicates and amplifies that of Kelly, using the UK-DALE dataset \cite{kelly2015uk} and proposing modifications to the architectures there defined. These modifications allow the introduction of high frequency features and the adaptability of the dimension of the autoencoder's latent space according to the appliance. An introduction to ANNs can be found at \cite{kingma2014adam}.

This work involves developments that are independent, although related to each other. On the one hand, the custom meter and the data collection system make up the low level component of the project.
On the other hand, the software or signal processing part involves appliance identification over the PLAID database \cite{gao2014plaid}, neural network models for disaggregation and the theoretical high level discussion of unsupervised approaches. 

The contributions of this work are listed as follows:
\begin{itemize}
    \item The collection of the first NILM oriented dataset in Uruguay (\cref{susec:nilmuy}).
    \item The study of high frequency features that yield a competitive appliance identification performance when used through a Random Forest classifier (\cref{sec:plaid}).
    \item The validation of the usefulness of the algorithms presented in \cite{Kelly2015}.
    \item The proposal and testing of a method for including high frequency features as input of ANN models along with the deeper autoencoder variant (\cref{sec:models}).
    \item The assesment of the generalization error of ANN based methods for NILM (\cref{sec:res}).
    \item A macro and general description of unsupervised and scalable NILM methods (\cref{sec:unsupervised}).
\end{itemize}

This paper is organized as follows.
In \cref{sec:meter} the custom meter is introduced along with its characteristics.
This meter is integrated into the data collection system described in \cref{sec:system}. 
In \cref{sec:data} the data preparation is explained, including synthetic data and training, validation and testing splits.
Then, a detailed study on high frequency features is described in \cref{sec:plaid}, that is used to create the multivariate time series to be used by the models.
A total of seven models are described in \cref{sec:models}, making emphasis on the training (\cref{sec:training}) and the model selection (\cref{sec:selection}) procedures.
Next, the evaluation scheme and the results are presented in \cref{sec:res}. 
Furthermore, a discussion on the formulation of unsupervised approaches is provided in \cref{sec:unsupervised}. 

\section{Data collection}
\subsection{Custom meter}\label{sec:meter}
\begin{figure}[htbp]
\centerline{\includegraphics[scale=0.3]{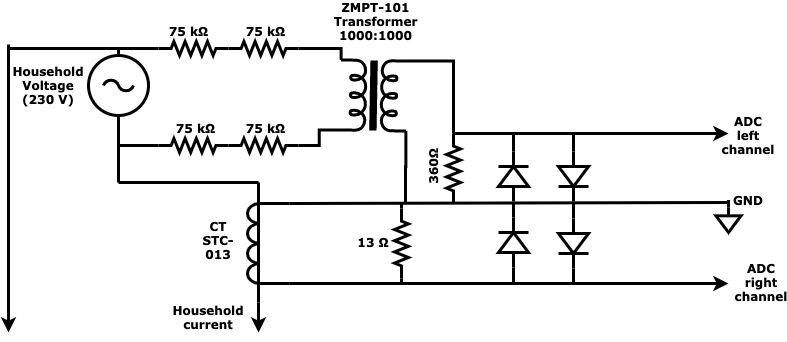}}
\caption{Circuit diagram used in the custom meter.}
\label{fig:nilm_circuit}
\end{figure}

The first step towards a NILM system is building a device that can measure at a high enough frequency.
The frequency requirements vary. For instance, public databases range from $1$ Hz to $44$ kHz, the private company Sense \cite{Sense} says its meter's sample rate is $1$MHz, and there exists applications of NILM using frequencies up to $100$ MHz \cite{patel2007flick}. 
Some experiments concerning the current clamp at use were carried out and revealed that it does not filter signal components under $7kHz$.
The spectral analysis of different appliances' current signals and the will to make comparison possible caused us to choose as Analog Digital Converter (ADC) a configurable soundcard named ``audio injector'' with a sample rate of up to $96$kHz, although the final sample rate was set to $14$kHz due to storage considerations. 
The meter was completed by printing a custom signal adaptation circuit, as shown in \cref{fig:nilm_circuit} and using a Raspberry Pi 3B+ that provides the required software flexibility and allows fast prototyping.

\subsection{Labeled data collection}\label{sec:system}
\begin{figure}[htbp]
\centerline{\includegraphics[scale=0.065]{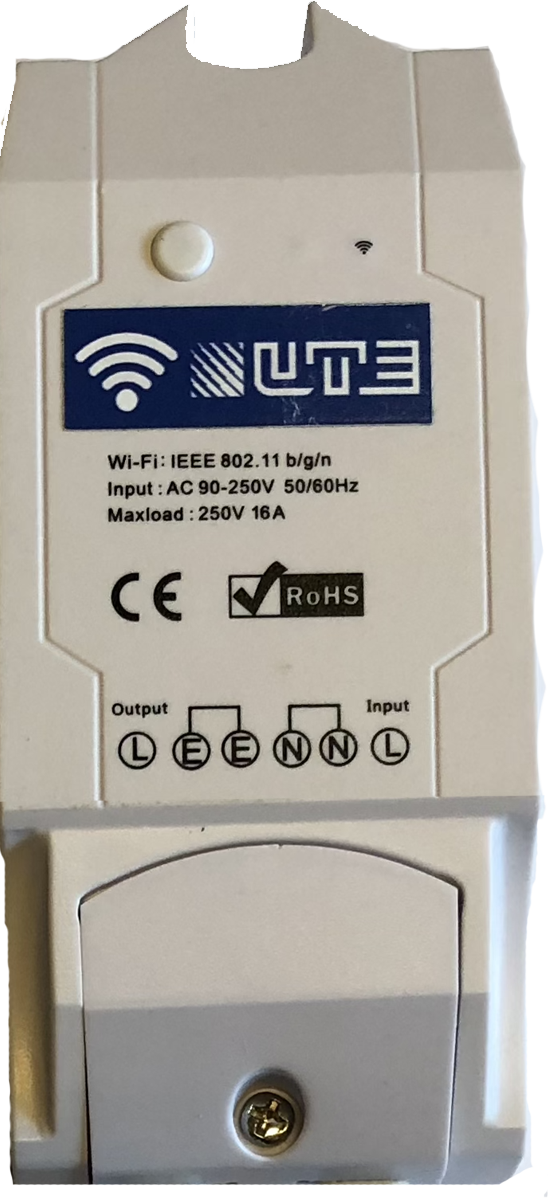}}
\caption{Intrusive meter provided by UTE.}
\label{fig:ute_meter}
\end{figure}

\begin{figure}[htbp]
\centerline{\includegraphics[scale=0.37]{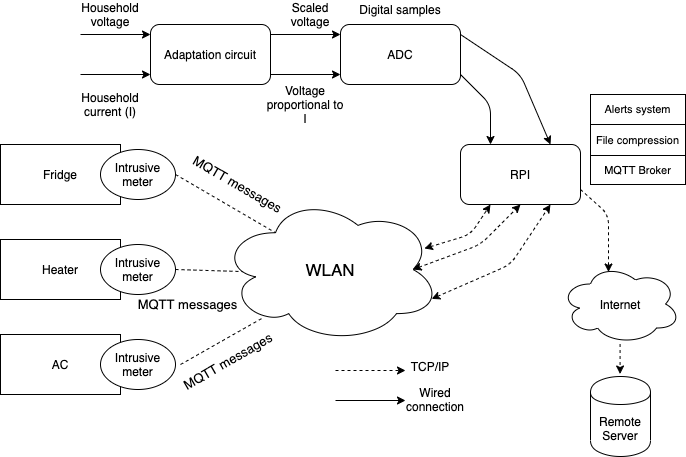}}
\caption{Complete diagram for the data collection system.}
\label{fig:nilm_diagram}
\end{figure}

In order to properly test any NILM system in terms of its disaggregation capabilities some labeled data set must be used. 
A data collector system should coordinate the non intrusive meter and some intrusive meters that measure individual appliances.
The system coordination was implemented on the non intrusive meter, more specifically, on the Raspberry Pi.
The computer's role was to save and compress the high frequency measurements in \verb|.flac| files while serving as a MQTT server to the intrusive meters.
The intrusive meters shown in \cref{fig:ute_meter} were provided by the Uruguayan national electricity utility (UTE) and report active power at a $1$ minute sample period.
The system was programmed to start running as soon it was energized, and it is able to send alerts if some component is malfunctioning.
The configuration files admit flexibility in terms of sample rate (up to $96$kHz), bit depth ($32$ bits), compression period ($1$ hour \verb|.flac|) and allow to write the data to external usb storage devices, save it in another computer in the same LAN or send it trough the TCP/IP to an external FTP server.
A complete visual description of the system is provided in \cref{fig:nilm_diagram}. 

\subsection{Uruguayan data}\label{susec:nilmuy}
Data were taken from two households in Montevideo, Uruguay, with a nominal frequency of $50$Hz and a nominal voltage of $230\text{V}_{\text{rms}}$. 
The total amount of recorded time sums up to $3$ months of data or $0.5$TB. 
In the first house $7$ intrusive meters were installed in the fridge, electric water heater, microwave, washing machine (from now on just ``washing''), air conditioner and bedroom plugs, whereas in the second house $8$ intrusive meters took the measurements from an electric oven, an electric water heater, two air conditioners, a fridge, a washing machine, a dishwasher and a kettle.
These appliances were the responsible for the majority of the households power consumption.
The next section presents the processing of these data that serve as input for the neural network based models.

\section{Data}\label{sec:data}
\subsection{Inputs and outputs}
The neural network algorithms to be presented in \cref{sec:models} belong to the class of supervised machine learning algorithms.
This means that the algorithm is obtained or trained from a set of values $(\textbf{X}, \textbf{Y})$ where $\textbf{X}$ is a vector containing all input examples and $\textbf{Y}$ is a vector containing their corresponding labels or target values.   
On the one hand, this work uses as input $\textbf{x} \in \textbf{X}$ a univariate power time series or a multivariate time series that includes the former in one of its dimensions.
These time series have a $6$ seconds period and their length is given by the size of a window that varies according to the appliance as shown in ~\cref{wsize}. 

\begin{table}[htbp]
\centering
\caption{Window size used.}
\begin{tabular}{|c|c|}
\hline
 & Window size (minutes) \\ \hline
Kettle  & 13                            \\ \hline
Fridge         & 60                            \\ \hline
Washing m. & 180                           \\ \hline
Microwave       & 10                            \\ \hline
Dishwasher     & 150                           \\ \hline
\end{tabular}
\label{wsize}
\end{table}

On the other hand, an output $\textbf{y} \in \textbf{Y}$ takes two possible forms, corresponding to the two ANN architectures to be presented in \cref{sec:models}.
The first of them involves three values: the beginning of the appliance activation, the end of the appliance activation, and the mean power consumed between these instants. 
An appliance's activation is extracted by a function that takes as parameters the minimum and maximum switched on time of the appliance, the on-power threshold and the border or padding for the window. 
The second possibility for $\textbf{y}$ is an univariate power time series of the appliance, which has the same length and period as the input. 

\subsection{Data preparation}
It should be noted that for Uruguayan data a first order hold is used to upsample the $1$ minute period measurements. 
The Uruguayan data was used only for evaluation and not for training. 
The training was based on the UK-DALE dataset, whose detailed description is found on \cite{kelly2015uk}. 
Succinctly, this database is formed up by measurements of five houses, three of which also include high frequency measurements. 
The set $\textbf{X}$ is built by activations extracted from this database together with non activations in equal proportion.
A non activation could be any window that does not fully include the functioning period of the appliance as defined by the activation-extracting function.

The multivariate time series for both Uruguayan and UK-DALE data had to be obtained. For the latter, web scrapping was used to download the $7.6$TB of data, from which the multivariate series values were computed for each required datetime. Comparison of power values extracted from the high frequency time series against UK-DALE's low frequency power data was made in order to check the correctness of the procedure. The code was reused on uruguayan data. The two high frequency features computed, namely form factor and phase shift of fundamental components of current and voltage, were selected for inclusion on the inputs after the analysis presented in the next section.

\subsection{Synthetic data}
Data augmentation was also effectuated by superposing to an appliance activation other appliances activations with some probability, defined as $p=0.4$ of adding a ``distractor'' appliance to the individual activation. The sum of these individual power values then composes the aggregate series to be used as input. For samples not containing activations of the target appliances, only ``distractor'' activations where included with probability $p$. It should be noted that synthetic data for multivariate series can not be constructed, as there are no individual measurements of the high frequency features.

\subsection{Dataset division}
The usual preparation of data in supervised learning algorithms involves dividing the dataset into three sets: training set, validation set, and testing or test set.
We follow this use, but there will be four test sets to be used.
For each appliance, measurements of one of the houses of the UK-DALE dataset are set aside for the test set I.
For the measurements in the other houses the last two weeks of data are also set aside for the test set II. The rest of the data is used to form the training and validation sets as will be shortly explained. 
The last two test sets correspond to the measurements taken from the two Uruguayan households.

From the time series that form the training and validation sets activations are extracted via the activation-extracting function. Also, non-activations are selected from the time period between two activations. The resulting activations dataset is approximately balanced, and it is split randomly into the training set ($80\%$) and the validation set ($20\%$).

\subsection{Preprocessing}
It is a well known practice the standardization and normalization of the signals to be used at training. The way we do that is extracting the mean standard deviation of the input training samples $\sigma_{\text{input}}$ and the maximum power output value $\text{max}_{\text{target}}$ of the training targets.
The preprocessing implies, for each input sample, independetly of which set it belongs to, its own mean is substracted before dividing by $\sigma_{\text{input}}$. For training, the targets are divided by $\text{max}_{\text{target}}$, and for prediction, the output is scaled for the same value.

\section{High Frequency Features}\label{sec:plaid}
In order to select the high frequency features to be included in the form of a multivariate time series together with the active power, a simpler problem was confronted.
This sub-problem is the identification of an appliance's name from its current and voltage signature. The PLAID database \cite{gao2014plaid} is made up of $1000+$ measurements of isolated appliances from $55$ different households at a $30$kHz sample rate. More than $30$ features were computed for each voltage and current waveform: power values as defined in \cite{6140203}, VI trajectory image, statistical moments, audio features, among others. As each instance contains the switching-on of the appliance, the transient was included in most of the instances.

\begin{figure}[htbp]
\centerline{\includegraphics[scale=0.125]{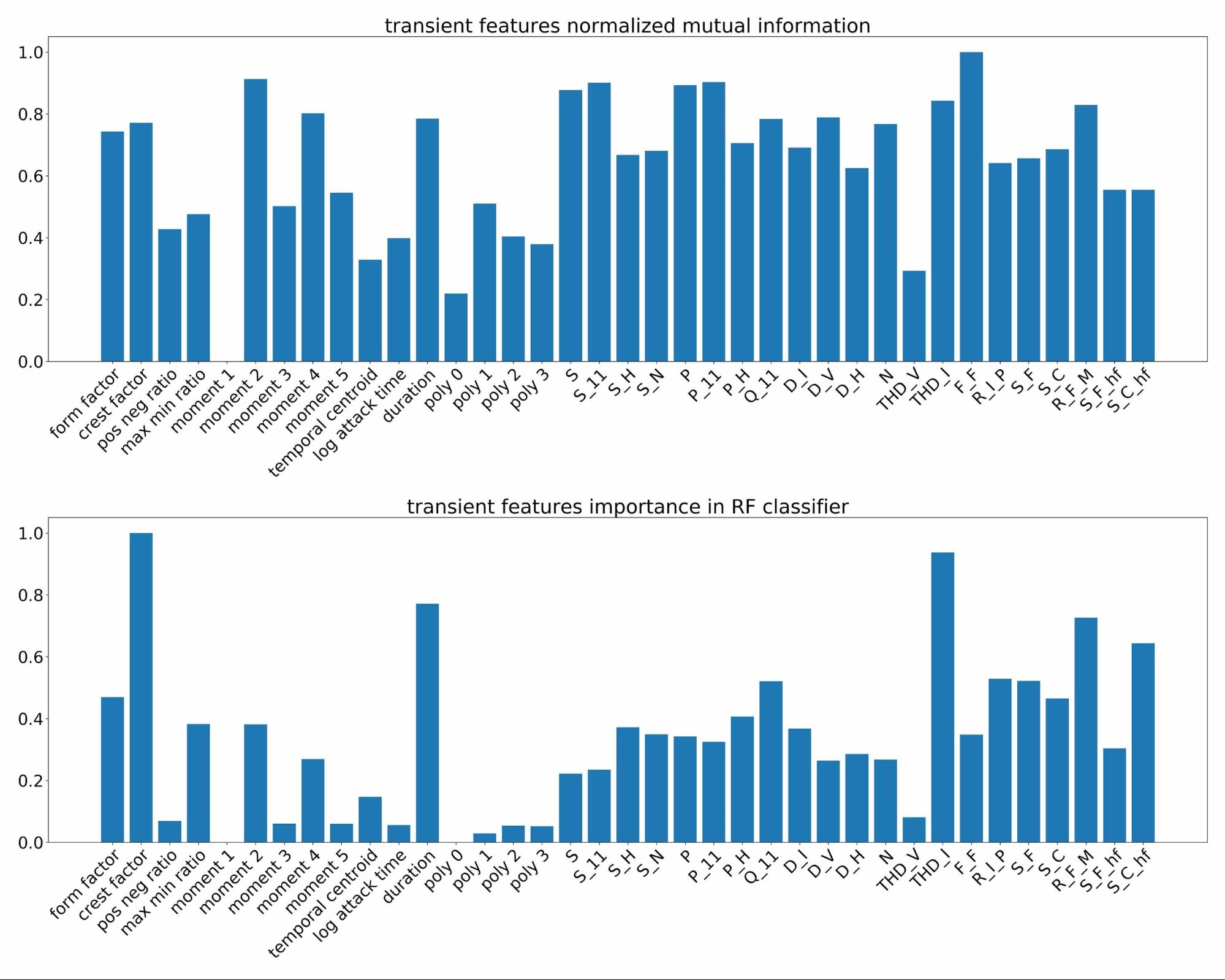}}
\caption{Transient features importance via random forest (top) and mutual information (bottom).}
\label{fig:mutual_info}
\end{figure}

Extracting this transient allowed to compute features over both transient and regime states. The features' importance was evaluated via Random Forest (RF) classifier and Mutual Information (MI) measure. ~\cref{fig:mutual_info} shows the normalized importance assigned to the transient features by the RF classifier and the MI measure. The criteria used to select the most important features was based on the assumption that any feature that is useful for the aggregate problem should be useful in this sub-problem too. The selected features comply with arbitrarily defined requirements: the selected features should be both transient and regime features (for instance, transient duration is not considered as a candidate) and should belong to the top $10$ features for transient and regime states under the importance criteria given by both the RF classifier and the MI measure. The two features that satisfy the requirements above are the form factor of the current and the phase shift between the fundamental component of the current and voltage signals.

The value of the original set of considered features is denoted by the high disaggregation performance obtained, showed in ~\cref{table:acc_std}, where 1-Nearest-Neighbor was used as a proxy classifier. It should be noted that the result that corresponds to all the possible features is only surpassed by the best result in \cite{sadeghianpourhamami2017comprehensive} by a $0.5$\% difference in accuracy, being superior to all other results found in the literature. This proves that the set of all features is powerful when used to feed a RF classifier.

\begin{table}
\centering
\caption{Performance over subsets of features.}
\begin{tabular}{|c|c|c|c|c|}
\hline
         {Instances} &
           \multicolumn{2}{c|}{1074} & \multicolumn{2}{c|}{1793} \\ \hline
\multicolumn{1}{|c|}{Features / Classifier}
&KNN&RF&KNN&RF\\ \hline
 \multicolumn{1}{|c|}{Transient} & 61.70 & 88.68$\pm$0.17& 59.35&87.06$\pm$0.06 \\ \hline
 \multicolumn{1}{|c|}{Steady state} & 75.88 & 87.24$\pm$0.28& 66.76&84.23$\pm$0.25 \\ \hline
 \multicolumn{1}{|c|}{Steady state + Transient} & -& 91.47$\pm$0.09& -&88.33$\pm$0.25 \\ \hline
 \multicolumn{1}{|c|}{Steady state + VI} & 75.97 & 86.67$\pm$0.49 &  66.82&84.14$\pm$0.43 \\ \hline
 \multicolumn{1}{|c|}{All features} & - & \textbf{92.79} $\pm$ 0.13& -&89.08$\pm$0.38 \\ \hline
 \multicolumn{5}{p{8cm}}{
 VI corresponds to pixels of the VI image. The tolerance is the standard deviation between the three runs.
}
\end{tabular}
\label{table:acc_std}
\end{table}

\section{Models}\label{sec:models}
\begin{figure}[htbp]
\centerline{\includegraphics[scale=0.18]{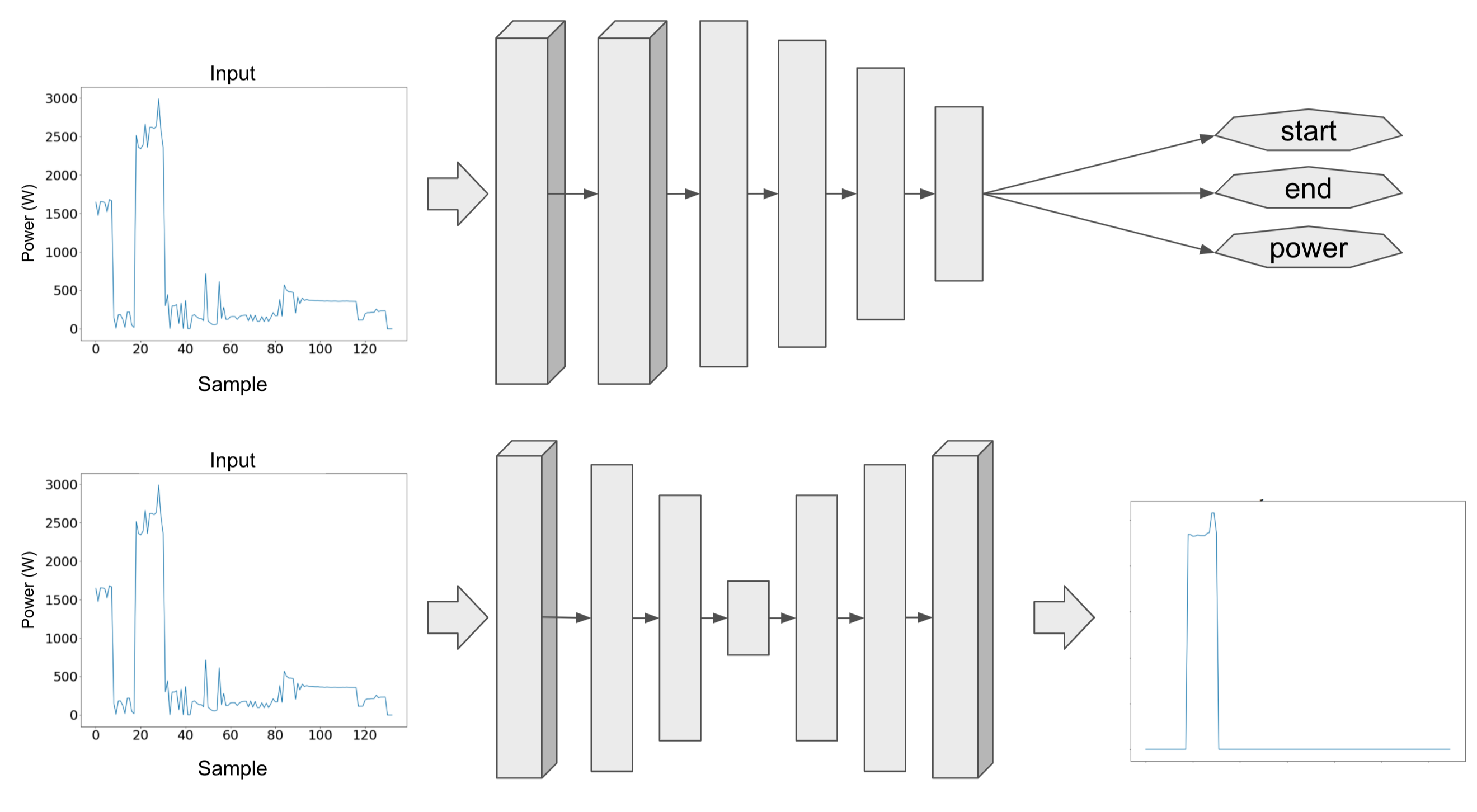}}
\caption{Diagram of the implemented ANN architectures. Rectangles network (top) and Autoencoder (bottom). 3D blocks correspond to convolution or upsampling layers.}
\label{fig:networks}
\end{figure}

The trained models are the originally used in \cite{Kelly2015} although some additional modifications were proposed. These models' names are ``autoencoder'' and ``rectangles'' as named in the previously cited work. The visual description is provided in ~\cref{fig:networks} and the Tensorflow model is exposed in the ~\cref{ap:tf_models}. The proposed modifications correspond to:
\begin{itemize}
    \item Changing the first convolutional layer so it is able to get mutivariate time series as input.
    \item Making the autoencoder deeper and the code length dependent on the input window size.
\end{itemize}
The seven models considered are presented in ~\cref{variants}. The ``baseline models'' are the ones trained with the augmented low frequency training set, i.e. the ones trained with synthetic data. 
\begin{table}[htbp]
\centering
\caption{Used models.}
\begin{tabular}{c|c|c|c|c|}
\cline{2-5}
                                  & Low freq & Synthetic data & High freq & "Big" \\ \hline
\multicolumn{1}{|c|}{Rectangles}   & Yes      & Yes            & Yes       & No    \\ \hline
\multicolumn{1}{|c|}{Autoencoder} & Yes      & Yes            & Yes       & Yes   \\ \hline
\end{tabular}
\label{variants}
\end{table}

\subsection{Training}\label{sec:training}
\begin{figure}[htbp]
\centerline{\includegraphics[scale=0.4]{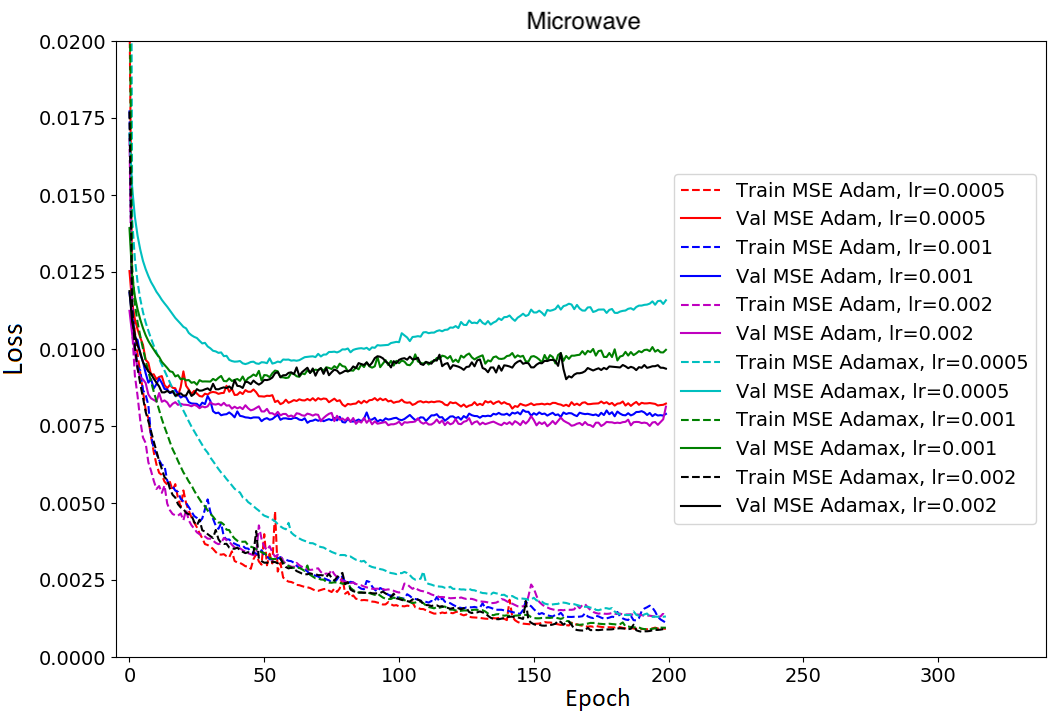}}
\caption{Losses evolution during training. Training loss (solid line). Validation loss (dashed line). The loss function is the MSE.}
\label{fig:micro_loss}
\end{figure}

It is common knowledge that ANNs must be trained. Any training procedure involves finding a set of weights that achieves a low loss over the training set while at the same time keeping the loss over the validation set controlled.
We found that the training delicate, as it is easy for the optimizers to get stuck into local minima, being the most evident case any set of parameters that yields always the same output.
To avoid local minima and get a good performance, multiple runs were made for each model. To find a good set of parameters for each model, two techniques were used.
The first involved a grid search in the space of optimizers. After discarding Adadelta \cite{zeiler2012adadelta} and large learning rates, six points were tried. 
These points arise from the combination of the learning rates values of $0.002, 0.001, 0.0005$ and the Adam   and Adamax optimizers \cite{kingma2014adam}.
The training procedure for one of the models for the microwave is shown in ~\cref{fig:micro_loss}.
The second technique involves tracking the error over the validation set and saving the model corresponding to the iteration with the lowest error.
The number of iterations for every model ran is $200$. This training procedure gives a total of $6$ runs per model $\times$ the $7$ models evaluated $\times$ $5$ appliances $=210$ runs. The $42000$ iterations were computed on a NVIDIA TITAN Xp graphics card. 

\begin{table}[htbp]
\centering
\caption{AUC of best experiments for the microwave.}
\begin{tabular}{|c|c|c|c|c|}
\hline
 & Low Freq & Synthetic Data & High Freq & ``Big'' \\ \hline
Rectangles & 0.933 & 0.937 & 0.927 & - \\ \hline
Autoencoder & 0.936 & 0.944 & \textbf{0.949} & 0.932 \\ \hline
\end{tabular}
\label{tab:auc_modelos}
\end{table}

\subsection{Selection}\label{sec:selection}
The goal is to get the ``best'' model for each appliance, i.e. to select for each appliance only one model of the $42$. As the loss function is the mean squared error between the output series and the target series it is not expected for this regression score to be unacceptably large for any model.
After manually analyzing the output we found that the Area Under the Curve (AUC) value of the Receiving Operating Characteristic (ROC) curve is strongly related to a visually good performing model. This metric will be used in the next steps to select the best models. 
For each of the $7$ models per appliance the selected set of parameters comes from choosing the weights that achieve the largest AUC between the $6$ sets of weights of the grid. 

To calculate the ROC the problem has to be turned into a classification one. 
To make this possible two criteria have to be defined: (i) the definition of the ground truth label, using for this the activation-extracting function described in \cref{sec:data}, and (ii) the definition of the predicted class, that is done by defining a threshold for the maximum on the predicted power series.
This first step gives tables similar to ~\cref{tab:auc_modelos} for each appliance. From these tables the best of the seven models is chosen by maximizing the AUC again. The results of which the best model is for every appliance are shown in ~\cref{sel_model}. Full results are included in the Appendix~\ref{ap:aucs}.

\begin{table}[htbp]
\centering
\caption{Selected model.}
\begin{tabular}{c|c|}
\cline{2-2}
                                      & Selected model             \\ \hline
\multicolumn{1}{|c|}{Kettle}          & High frequency autoencoder \\ \hline
\multicolumn{1}{|c|}{Fridge}          & High frequency rectangles   \\ \hline
\multicolumn{1}{|c|}{Washing m.} & High frequency rectangles   \\ \hline
\multicolumn{1}{|c|}{Microwave}       & High frequency autoencoder \\ \hline
\multicolumn{1}{|c|}{Dish washer}     & Big autoencoder            \\ \hline
\end{tabular}
\label{sel_model}
\end{table}

\section{Results}\label{sec:res}
Once the models are chosen, there are many evaluations to be made. These arise from the combination of two evaluation procedures, ``rolling window'' and ``activations''. These procedures will be described in ~\cref{sec:procedures}.
Besides the evaluation procedures there are the four test sets described in ~\cref{sec:data}.
Combining the different test sets with the two evaluation procedures allows finding the answers to the following questions:
\begin{enumerate}
    \item How do the models work for what they were trained for? - this corresponds to the evaluation over test set II using the ``activations'' procedure.\label{q:q1}
    \item How do the models generalize for unseen appliances? - this corresponds to the evaluation over test set I using the ``activations'' procedure.\label{q:q2}
    \item How would the models behave in a real case scenario? - this corresponds to the evaluation over test set I using the ``rolling window'' procedure.\label{q:q3}
\end{enumerate}

Models are further compared with the performance of the best models found at \cite{Kelly2015} as a reference. These are the ones that include synthetic data into the training set.
Finally, when answering the last question the value of the metrics to be introduced in  the next subsection is reported.

\subsection{Metrics}\label{sec:metrics}
To report the results a few common metrics were selected, in line with \cite{Kelly2015}. The metrics for the regression problem quantify how well the outputs approximate the targets, and are affected when time shifts occur. The classification metrics are agnostic to in-window time shifts, and only consider if the appliance is detected as energized or not. The regression metrics to be presented are the Mean Absolute Error (MAE) and Relative Error In Total Energy (REITE):
\begin{equation}
    \text{REITE} = \frac{|\hat{E} - E|}{\max{(\hat{E}, E)}}
\end{equation}
\begin{equation}
    \text{MAE} = \frac{1}{N} \sum_{t}{|\hat{y}_t - y_t|}
\end{equation}
where $N$ is the total number of power values considered. The predictions can be Positive (P) or Negative (N) and result True (T) if they are equal to the ground truth label or False (F) otherwise. Traditional classification metrics are defined as:
\begin{equation}
    \text{recall} = \frac{\text{TP}}{\text{TP} + \text{FN}}
\end{equation}

\begin{equation}
    \text{precision} = \frac{\text{TP}}{\text{TP} + \text{FP}}
\end{equation}

\begin{equation}
    \text{accuracy} = \frac{\text{TP} + \text{TN}}{\text{TP} + \text{TN} + \text{FP} + \text{FN}}
\end{equation}

\begin{equation}
    \text{F}_1 = 2 \cdot \frac{\text{precision} \cdot \text{recall}}{\text{precision}+\text{recall}}
\end{equation}

\subsection{Evaluation procedures}\label{sec:procedures}
Two evaluation procedures are used in this work. The ``activations'' procedure involves extracting activations and non-activations from a test set, making estimates for each window and comparing that with the ground truth.
It is the simplest evaluation procedure and it has two important features: (i) the resulting windows are approximately balanced and (ii) the method is the same as the one used over the validation set.

Alternatively, the ``rolling window'' procedure is as unbalanced as the real use of the appliances. Furthermore, it is applied over the whole time series, resembling a real use case.
This procedure starts estimating the output for each one of the inputs determined by a rolling window with stride$=1$. 
This is, estimate one output, shift the window $6$ seconds, estimate another output, and so on.
At the end, for every datetime, there are $w=\verb|windowsize|$ estimates, that are averaged and multiplied by a factor of $\frac{w}{w-2a}$, where $a$ is the average activation length for the appliance calculated over the training and validation sets, in order to account for the fact that our ANNs only recognize \emph{complete} activations.
After the estimation stage, the same time series is divided in non overlapping windows over which the resulting estimates are compared with the ground truth.

\subsection{Results}
Numerical results will be summarized in this sub-section. First, it can be seen in ~\cref{tab:act_todo} that the best models yields the best AUC for three of the five appliances for the case depicted in question \ref{q:q1}.
This means that the models perform very well when evaluated over new instances of previously seen appliances and that the training and selection procedures were correctly designed.
Second, ~\cref{tab:act_todo} shows that the performance of the best models declines more than the performance of the reference models when testing over unseen appliances. The conclusion is that the generalization capability of the model to probability distributions other than the one generating the training and validation data is limited.
Finally, Tables \ref{tab:rolling} and \ref{tab:metricas_no_visto} summarizes the performance obtained on the real case scenario over the unseen appliances of the UK-DALE database. A MAE under $60$W for almost every appliance and good classification metrics for three of five appliances are indicative of acceptable results. Note here that the threshold used to get the classification scores was defined for each appliance by maximizing the $\text{F}_1$ score over the validation set.

\begin{table}[htbp]
\centering
\caption{AUCs scores via activations methodology.}
\begin{tabular}{cc|c|c|}
\cline{3-4} 
 &  & Test set II & Validation set \\ \hline
\multicolumn{1}{|c|}{\multirow{3}{*}{Kettle}} & Best model & \underline{\textbf{0.999}} & \textbf{0.984} \\ \cline{2-4}  \multicolumn{1}{|c|}{} & Autoencoder & 0.959 & \underline{0.977} \\ \cline{2-4}  \multicolumn{1}{|c|}{} & Rectangles & 0.941 & \underline{0.982} \\ \hline \hline
\multicolumn{1}{|c|}{\multirow{3}{*}{Fridge}} & Best model & \textbf{\underline{0.941}} & \textbf{0.901} \\ \cline{2-4} 
\multicolumn{1}{|c|}{} & Autoencoder & 0.500 & 0.500 \\ \cline{2-4} 
\multicolumn{1}{|c|}{} & Rectangles & \underline{0.871} & 0.811 \\ \hline \hline
\multicolumn{1}{|c|}{\multirow{3}{*}{Washing m.}} & Best model & 0.850 & \underline{\textbf{0.951}} \\ \cline{2-4} 
\multicolumn{1}{|c|}{} & Autoencoder & \textbf{\underline{0.884}} & 0.875 \\ \cline{2-4} 
\multicolumn{1}{|c|}{} & Rectangles & 0.812 & \underline{0.900} \\ \hline \hline
\multicolumn{1}{|c|}{\multirow{3}{*}{Microwave}} & Best model & \underline{0.976} & \textbf{0.949} \\ \cline{2-4} 
\multicolumn{1}{|c|}{} & Autoencoder & 0.932 & \underline{0.944} \\ \cline{2-4} 
\multicolumn{1}{|c|}{} & Rectangles & \underline{0.976} & 0.937 \\ \hline \hline
\multicolumn{1}{|c|}{\multirow{3}{*}{Dishwasher}} & Best model & 0.947 & \underline{\textbf{0.997}} \\ \cline{2-4} 
\multicolumn{1}{|c|}{} & Autoencoder & \textbf{0.986} & \underline{0.989} \\ \cline{2-4} 
\multicolumn{1}{|c|}{} & Rectangles & 0.954 & \underline{0.981} \\ \hline
\multicolumn{4}{p{7cm}}{Evaluation via ``activations'' methodology. Bold font indicates the best score achieved by the three models. Underlines indicate best score achieved between datasets for each model.}
\end{tabular}
\label{tab:testvsval}
\end{table}

\begin{table}[htbp]
\caption{AUC scores via activations methodology.}
\centering
\begin{tabular}{cc|c|c|}
 \cline{3-4} 
 &  & Test set II & Test set I \\ \hline
\multicolumn{1}{|c|}{\multirow{3}{*}{Kettle}} & Best model & \underline{\textbf{0.999}} & 0.983 \\ \cline{2-4} 
\multicolumn{1}{|c|}{} & Autoencoder & 0.959 & \underline{0.993} \\ \cline{2-4} 
\multicolumn{1}{|c|}{} & Rectangles & 0.941 & \underline{\textbf{0.994}} \\ \hline \hline
\multicolumn{1}{|c|}{\multirow{3}{*}{Fridge}} & Best model & \underline{\textbf{0.941}} & 0.703 \\ \cline{2-4} 
\multicolumn{1}{|c|}{} & Autoencoder & 0.500 & 0.500 \\ \cline{2-4} 
\multicolumn{1}{|c|}{} & Rectangles & 0.871 & \underline{\textbf{0.887}} \\ \hline \hline
\multicolumn{1}{|c|}{\multirow{3}{*}{Washing m.}} & Best model & \underline{0.850} & 0.795 \\ \cline{2-4} 
\multicolumn{1}{|c|}{} & Autoencoder & \textbf{0.884} & \underline{\textbf{0.908}} \\ \cline{2-4} 
\multicolumn{1}{|c|}{} & Rectangles & 0.812 & \underline{0.884} \\ \hline \hline
\multicolumn{1}{|c|}{\multirow{3}{*}{Microwave}} & Best model & \underline{0.976} & 0.879 \\ \cline{2-4} 
\multicolumn{1}{|c|}{} & Autoencoder & \underline{0.932} & 0.891 \\ \cline{2-4} 
\multicolumn{1}{|c|}{} & Rectangles & \underline{0.976} & \textbf{0.962} \\ \hline \hline
\multicolumn{1}{|c|}{\multirow{3}{*}{Dishwasher}} & Best model & 0.947 & \underline{0.962} \\ \cline{2-4} 
\multicolumn{1}{|c|}{} & Autoencoder & \underline{\textbf{0.986}} & \textbf{0.984} \\ \cline{2-4} 
\multicolumn{1}{|c|}{} & Rectangles & 0.954 & \underline{0.973} \\ \hline
\multicolumn{4}{p{7cm}}{Evaluation via ``activations'' methodology. Bold font indicates the best score achieved by the three models. Underlines indicate best score achieved between datasets for each model.}
\end{tabular}
\label{tab:act_todo}
\end{table}

\begin{table}[htbp]
\label{tab:rolling}
\centering
\caption{AUCs scores via rolling window methodology.}
\begin{tabular}{c|c|c|}
\cline{2-3}
                                 & Test set II & Test set I \\ \hline
\multicolumn{1}{|c|}{Kettle}     & 1.000       & 0.998      \\ \hline
\multicolumn{1}{|c|}{Fridge}     & 0.854       & 0.751      \\ \hline
\multicolumn{1}{|c|}{Washing m.} & 0.763       & 0.864      \\ \hline
\multicolumn{1}{|c|}{Microwave}  & 0.973       & 0.956      \\ \hline
\multicolumn{1}{|c|}{Dishwasher} & 0.898       & 0.962      \\ 
\hline
\end{tabular}
\end{table}

\begin{table}[htbp]
\label{tab:metricas_no_visto}
\centering
\caption{Results over Test Set I of the best models via rolling window methodology.}
\begin{tabular}{c|c|c|c|c|c|c|}
\cline{2-7}
                                      & Acc. & Prec. & Recall & F1 & MAE    & REITE \\ \hline
\multicolumn{1}{|c|}{Kettle}          & 0.987    & 0.686     & 0.967  & 0.802    & 22.48  & 0.609 \\ \hline
\multicolumn{1}{|c|}{Fridge}          & 0.585    & 0.545     & 0.959  & 0.695    & 42.04  & 0.305 \\ \hline
\multicolumn{1}{|c|}{Washing m..} & 0.612    & 0.107     & 0.904  & 0.191    & 237.98 & 0.962 \\ \hline
\multicolumn{1}{|c|}{Microwave}       & 0.835    & 0.019     & 0.964  & 0.038    & 58.48  & 0.173 \\ \hline
\multicolumn{1}{|c|}{Dish washer}     & 0.961    & 0.679     & 0.743  & 0.710    & 45.00  & 0.639 \\ \hline
\end{tabular}
\end{table}

The results obtained for the Uruguayan households are presented in \cref{tab:uruguay}. For the household that only contains three appliances the performance is bad. No model generalizes well to this household.
For the other one, a better performance was obtained.
The models could be useful to get some information for three of the five appliances, namely the  ones in which $\text{F}_1$ score is high. However, these models' performance is far from excellent and they do not seem suitable for standalone use. 

We present visual examples of some of the experiments under the ``rolling window'' evaluation scheme. The softness of the curve is due to the averaging needed on the estimation stage.

\begin{table*}[t]
\caption{Results over Uruguayan households via rolling window methodology.}
\begin{tabular}{c|c|c|c|c|c|c|c|c|c|c|c|c|}
\cline{2-13}
 & \multicolumn{6}{c|}{Household 1} & \multicolumn{6}{c|}{Household 2} \\ \hline
\multicolumn{1}{|c|}{Appliance} & Accuracy & Precision & Recall & F1 & MAE (W) & REITE & Accuracy & Precision & Recall & F1 & MAE (W) & REITE \\ \hline
\multicolumn{1}{|c|}{Kettle} & - & - & - & - & - & - & 0.953 & 0.286 & 0.545 & 0.375 & 75 & 0.909 \\ \hline
\multicolumn{1}{|c|}{Fridge} & 0.759 & 0.781 & 0.959 & 0.861 & 143 & 0.507 & 0.918 & 0.990 & 0.926 & 0.957 & 96 & 0.071 \\ \hline
\multicolumn{1}{|c|}{Washing m.} & 0.071 & 0.057 & 1.000 & 0.107 & 793 & 0.995 & 0.323 & 0.300 & 1.000 & 0.462 & 691 & 0.971 \\ \hline
\multicolumn{1}{|c|}{Microwave} & 0.506 & 0.018 & 0.650 & 0.036 & 71 & 0.920 & 0.533 & 0.066 & 0.818 & 0.122 & 90 & 0.880 \\ \hline
\multicolumn{1}{|c|}{Dish washer} & - & - & - & - & - & - & 0.859 & 0.720 & 0.720 & 0.720 & 150 & 0.478 \\ \hline
\end{tabular}
\label{tab:uruguay}
\end{table*}

\section{About unsupervised approaches}\label{sec:unsupervised}

ANNs are being used widely in the supervised setting, and they present a good performance when carefully trained with data that resembles the real case scenario.
But the best possible performance of supervised methods is bounded by the characteristics of the training datasets.
These have to be extensive and include many kind of appliances and their combinations in order to make possible an adequate generalization capability. Synthesizing data is a good approach to solve this issue. Another approach, that we will present here as a theoretical macro design, is to shift from the supervised approach to an unsupervised learning approach. Notwithstanding, there are other methods that involve learning and could be useful, such as active learning approaches. 

The unsupervised approach is presented in a generic way. This takes into account the requirements of a NILM algorithm and focuses on its scaling ability.
Any learning algorithm should have a loss function or error signal.
The most intuitive error definition for the unsupervised case is the reconstruction error of the aggregate power time series. The minimization of this error implies a good estimation of the aggregate power signal.
Furthermore, the estimation of the aggregate power should be done from useful data, for instance from the appliances' power consumption estimates. The name we assign to the block that creates an estimated aggregate signal from useful information is ``simulator''. What this framing is implying is that minimizing the error between the aggregate power signal and an estimation of it from some useful estimates is forcing the useful estimates to be accurate.
The last block needed to complete this generic formulation is the black box that computes the useful estimates, which represents the disaggregation algorithm that adjusts its parameters according to the error signal.

\begin{figure}[htbp]
\centerline{\includegraphics[scale=0.3]{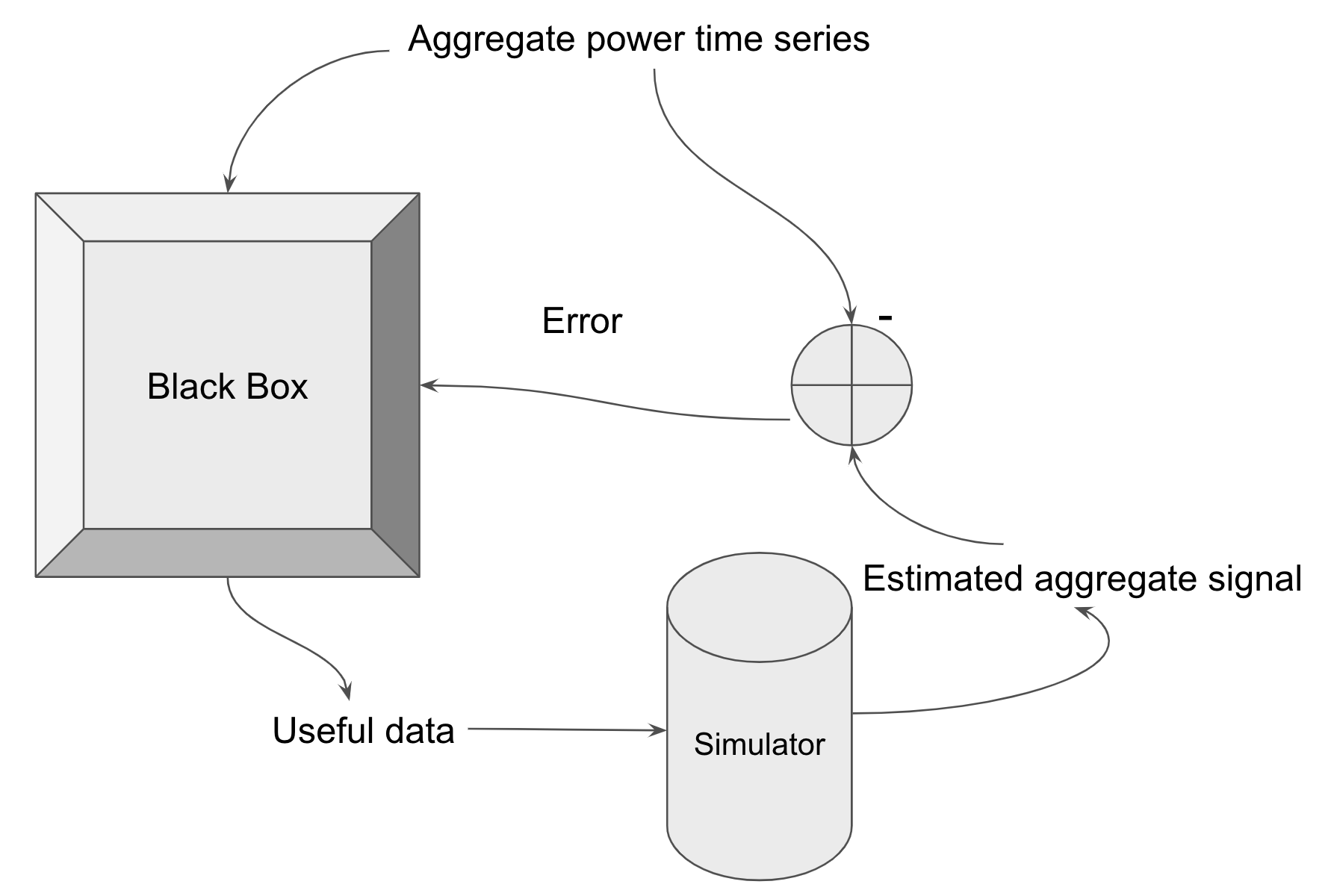}}
\caption{Diagram of the general unsupervised NILM algorithm.}
\label{fig:black_box}
\end{figure}

Every component of the proposed general algorithm is presented in \cref{fig:black_box} and is a subject of research. 
We consider that most of the unsupervised approaches can be framed this way, although the presented building blocks are not necessarily simple nor independent.

\section{Conclusions}\label{sec:conclusions}

To conclude, let us note that a NILM system can be built and evaluated as was done in this work. The implemented data collector system is robust, as it was capable of collect measurements without intervention for three months.

The first conclusion refers to the high frequency features. These were studied and using the MI measure and the RF classifier two of them were selected: phase shift and form factor. The proposed methodology to integrate high frequency information in the algorithm, that involved obtaining a low frequency multivariate time series of descriptors, was correctly implemented. Furthermore, the model selection procedure selected the models that included high frequency information as the best model for almost all studied appliances. The good performance obtained over the test set II, the one containing known appliances, proves the value added by these high frequency features.

Secondly, neural network based models can achieve very good performance metrics for appliances that were seen during training. This is the expected behavior of correctly trained supervised approaches. 

Notwithstanding, the generalization power to appliances unseen during training is limited, although it can not be ignored that the number of houses used for training is less than $5$. The results are not as good over the test set I than over the test set II. This performance decline is more notorious for test sets corresponding to the Uruguayan households. However, for one of these households, the AUC values surpass $0.8$, denoting a respectable performance.

Moreover, the AUC was used as the main metric for model selection, using the ROC as was recommended in \cite{zeifman2011nonintrusive}. Finally, unsupervised approaches were decomposed in a few building blocks, and we hope this helps conceptualize this kinds of approaches in the future.

\section{Acknowledgements}
The authors thankfully acknowledge the financial support provided by Uruguay’s National Research and Innovation Agency (ANII), the Julio Ricaldoni Foundation and the National Administration of Power Plants and Electrical Transmissions. 

\bibliographystyle{plain}
\bibliography{refs}

\newpage
\appendix 

\label{ap:tf_models}
\begin{figure}[ht]
    \centering
    \footnotesize
    \begin{verbatim}
________________________________________________
Layer (type)        Output Shape        Param #   
================================================
conv1d_2 (Conv1D)   (None, 127, 8)      40        
________________________________________________
flatten_1 (Flatten) (None, 1016)        0         
________________________________________________
dense_3 (Dense)     (None, 1016)        1033272   
________________________________________________
dense_4 (Dense)     (None, 128)         130176    
________________________________________________
dense_5 (Dense)     (None, 1016)        131064    
________________________________________________
reshape_1 (Reshape) (None, 127, 8)      0         
________________________________________________
zero_padding1d_1    (None, 130, 8)      0         
________________________________________________
conv1d_3 (Conv1D)   (None, 130, 1)      33        
================================================
Total params: 1,294,585
Trainable params: 1,294,585
Non-trainable params: 0
\end{verbatim}
    \caption{Autoencoder for kettle (window length 130).}
    \label{fig:denoising_autoencoder}
\end{figure}

\begin{figure}[h]
    \centering
    \footnotesize
        \begin{verbatim}
________________________________________________
Layer (type)        Output Shape        Param #   
================================================
conv1d_4 (Conv1D)   (None, 127, 16)     80        
________________________________________________
conv1d_5 (Conv1D)   (None, 124, 16)     1040      
________________________________________________
flatten_2 (Flatten) (None, 1984)        0         
________________________________________________
dense_6 (Dense)     (None, 4096)        8130560   
________________________________________________
dense_7 (Dense)     (None, 3072)        12585984  
________________________________________________
dense_8 (Dense)     (None, 2048)        6293504   
________________________________________________
dense_9 (Dense)     (None, 512)         1049088   
________________________________________________
dense_10 (Dense)    (None, 3)           1539      
================================================
Total params: 28,061,795
Trainable params: 28,061,795
Non-trainable params: 0
        \end{verbatim}
    \caption{Rectangles network for kettle.}
    \label{fig:arquitectura_rectangulos}
\end{figure}

\newpage
\begin{figure}[ht]
    \centering
    \footnotesize
        \begin{verbatim}
________________________________________________
Layer (type)        Output Shape        Param #
================================================
conv1d_2 (Conv1D)   (None, 127, 16)     208
________________________________________________
conv1d_3 (Conv1D)   (None, 124, 16)     1040
________________________________________________
flatten_1 (Flatten) (None, 1984)        0
________________________________________________
dense_3 (Dense)     (None, 4096)        8130560
________________________________________________
dense_4 (Dense)     (None, 3072)        12585984
________________________________________________
dense_5 (Dense)     (None, 2048)        6293504
________________________________________________
dense_6 (Dense)     (None, 512)         1049088
________________________________________________
dense_7 (Dense)     (None, 3)           1539
================================================
Total params: 28,061,923
Trainable params: 28,061,923
Non-trainable params: 0
        \end{verbatim}
    \caption{High frequency rectangles network for kettle.}
    \label{fig:arquitectura_rectangulos_hf}
\end{figure}

\begin{figure}[h]
    \centering
    \footnotesize
        \begin{verbatim}
________________________________________________
Layer (type)        Output Shape        Param #
================================================
conv1d (Conv1D)     (None, 127, 8)      104
________________________________________________
flatten (Flatten)   (None, 1016)        0
________________________________________________
dense (Dense)       (None, 1016)        1033272
________________________________________________
dense_1 (Dense)     (None, 128)         130176
________________________________________________
dense_2 (Dense)     (None, 1016)        131064
________________________________________________
reshape (Reshape)   (None, 127, 8)      0
________________________________________________
zero_padding1d      (None, 130, 8)      0
________________________________________________
conv1d_1 (Conv1D)   (None, 130, 1)      33
================================================
Total params: 1,294,649
Trainable params: 1,294,649
Non-trainable params: 0
        \end{verbatim}
    \caption{High frequency autoencoder for kettle.}
    \label{fig:ae_hf}
\end{figure}

\newpage
\begin{figure}[ht]
    \centering
    \footnotesize
        \begin{verbatim}
________________________________________________
Layer (type)        Output Shape        Param #   
================================================
conv1d (Conv1D)     (None, 127, 8)      40        
________________________________________________
conv1d_1 (Conv1D)   (None, 124, 8)      264       
________________________________________________
flatten (Flatten)   (None, 992)         0         
________________________________________________
dense (Dense)       (None, 1016)        1008888   
________________________________________________
dense_1 (Dense)     (None, 254)         258318    
________________________________________________
dense_2 (Dense)     (None, 13)          3315      
________________________________________________
dense_3 (Dense)     (None, 254)         3556      
________________________________________________
dense_4 (Dense)     (None, 1016)        259080    
________________________________________________
reshape (Reshape)   (None, 127, 8)  0         
________________________________________________
zero_padding1d      (None, 130, 8)  0         
________________________________________________
conv1d_2 (Conv1D)   (None, 130, 1)  33        
================================================
Total params: 1,533,494
Trainable params: 1,533,494
Non-trainable params: 0
        \end{verbatim}
    \caption{``Big'' autoencoder for kettle.}
    \label{fig:big_auto}
\end{figure}

\label{ap:aucs}

\onecolumn
\begin{sidewaystable}
\caption{Validation AUC for all trained models.}
\begin{tabular}{|l|l|l|l|l|l|l|l|l|l|l|}
\hline
Elec & Model & Freq & Run & Adam .0005 & Adam 0.001 & Adam 0.002 & AMax 0.0005 & AMax 0.001 & AMax 0.002 & Best \\ \hline
dish & rectangulos & data & lf\_0 & 0.9863 & 0.9914 & 0.9692 & 0.9875 & 0.9881 & 0.977 & 0.9914 \\ \hline
dish & rectangulos & data & lf\_syn\_0 & 0.9806 & 0.9794 & 0.9775 & 0.9777 & 0.9753 & 0.9723 & 0.9806 \\ \hline
dish & rectangulos & data\_hf & hf\_0 & 0.9778 & 0.9739 & 0.9708 & 0.9893 & 0.9825 & 0.9491 & 0.9893 \\ \hline
dish & autoencoder & data & ae\_lf\_run & 0.9763 & 0.5 & 0.5 & 0.9846 & 0.9793 & 0.5 & 0.9846 \\ \hline
dish & autoencoder & data & ae\_big\_lf\_run & 0.9752 & 0.5 & 0.5 & 0.9811 & 0.5 & 0.9968 & 0.9968 \\ \hline
dish & autoencoder & data & ae\_lf\_syn\_run & 0.9752 & 0.9799 & 0.5 & 0.9886 & 0.9737 & 0.5 & 0.9886 \\ \hline
dish & autoencoder & data\_hf & ae\_hf\_run & 0.977 & 0.9753 & 0.5 & 0.9412 & 0.9661 & 0.5 & 0.977 \\ \hline
Fridge & rectangulos & data & lf\_0 & 0.8001 & 0.5 & 0.5 & 0.8102 & 0.7953 & 0.8139 & 0.8139 \\ \hline
fridge & rectangulos & data & lf\_syn\_0 & 0.811 & 0.5 & 0.5 & 0.8037 & 0.809 & 0.8028 & 0.811 \\ \hline
fridge & rectangulos & data\_hf & hf\_0 & 0.8689 & 0.6374 & 0.5 & 0.9012 & 0.868 & 0.8797 & 0.9012 \\ \hline
fridge & autoencoder & data & ae\_lf\_run & 0.5 & 0.5 & 0.5 & 0.5 & 0.5 & 0.5 & 0.5 \\ \hline
fridge & autoencoder & data & ae\_big\_lf\_run & 0.5 & 0.5 & 0.5 & 0.5 & 0.5 & 0.5 & 0.5 \\ \hline
fridge & autoencoder & data & ae\_lf\_syn\_run & 0.5 & 0.5 & 0.5 & 0.5 & 0.5 & 0.5001 & 0.5001 \\ \hline
fridge & autoencoder & data\_hf & ae\_hf\_run & 0.5 & 0.5 & 0.5 & 0.8312 & 0.4998 & 0.5 & 0.8312 \\ \hline
kettle & rectangulos & data & lf\_0 & 0.9748 & 0.9767 & 0.5003 & 0.9707 & 0.9741 & 0.9686 & 0.9767 \\ \hline
kettle & rectangulos & data & lf\_syn\_0 & 0.9695 & 0.9757 & 0.5 & 0.9783 & 0.9759 & 0.9815 & 0.9815 \\ \hline
kettle & rectangulos & data\_hf & hf\_0 & 0.9652 & 0.9714 & 0.9529 & 0.9642 & 0.9673 & 0.9555 & 0.9714 \\ \hline
kettle & autoencoder & data & ae\_lf\_run & 0.9778 & 0.9769 & 0.9792 & 0.9791 & 0.9772 & 0.9781 & 0.9792 \\ \hline
kettle & autoencoder & data & ae\_big\_lf\_run & 0.9754 & 0.9726 & 0.9646 & 0.9779 & 0.9795 & 0.971 & 0.9795 \\ \hline
kettle & autoencoder & data & ae\_lf\_syn\_run & 0.9772 & 0.9768 & 0.9745 & 0.9768 & 0.9765 & 0.9755 & 0.9772 \\ \hline
kettle & autoencoder & data\_hf & ae\_hf\_run & 0.9763 & 0.984 & 0.979 & 0.9762 & 0.9759 & 0.9794 & 0.984 \\ \hline
microwave & rectangulos & data & lf\_0 & 0.8999 & 0.9205 & 0.5 & 0.9277 & 0.9331 & 0.9231 & 0.9331 \\ \hline
microwave & rectangulos & data & lf\_syn\_0 & 0.9125 & 0.9329 & 0.5221 & 0.9357 & 0.9368 & 0.912 & 0.9368 \\ \hline
microwave & rectangulos & data\_hf & hf\_0 & 0.9203 & 0.919 & 0.5007 & 0.9267 & 0.9247 & 0.9122 & 0.9267 \\ \hline
microwave & autoencoder & data & ae\_lf\_run & 0.934 & 0.9315 & 0.9256 & 0.9358 & 0.9339 & 0.9263 & 0.9358 \\ \hline
microwave & autoencoder & data & ae\_big\_lf\_run & 0.9126 & 0.9226 & 0.5 & 0.9321 & 0.932 & 0.9233 & 0.9321 \\ \hline
microwave & autoencoder & data & ae\_lf\_syn\_run & 0.9316 & 0.9337 & 0.934 & 0.9439 & 0.9318 & 0.9314 & 0.9439 \\ \hline
microwave & autoencoder & data\_hf & ae\_hf\_run & 0.942 & 0.9492 & 0.9374 & 0.5 & 0.9442 & 0.9407 & 0.9492 \\ \hline
washing & rectangulos & data & lf\_0 & 0.883 & 0.9002 & 0.8817 & 0.8893 & 0.8582 & 0.9037 & 0.9037 \\ \hline
washing & rectangulos & data & lf\_syn\_0 & 0.8823 & 0.8658 & 0.5 & 0.9004 & 0.8867 & 0.8919 & 0.9004 \\ \hline
washing & rectangulos & data\_hf & hf\_0 & 0.9252 & 0.9176 & 0.8893 & 0.9509 & 0.9419 & 0.9362 & 0.9509 \\ \hline
washing & autoencoder & data & ae\_lf\_run & 0.8774 & 0.8709 & 0.5 & 0.8774 & 0.5 & 0.5 & 0.8774 \\ \hline
washing & autoencoder & data & ae\_big\_lf\_run & 0.5 & 0.5 & 0.5 & 0.8496 & 0.5 & 0.5 & 0.8496 \\ \hline
washing & autoencoder & data & ae\_lf\_syn\_run & 0.8633 & 0.8746 & 0.5 & 0.8682 & 0.4983 & 0.5 & 0.8746 \\ \hline
washing & autoencoder & data\_hf & ae\_hf\_run & 0.8761 & 0.5 & 0.5 & 0.49 & 0.494 & 0.5 & 0.8761 \\ \hline
\end{tabular}
\label{table:auc_validacion}
\end{sidewaystable}

\end{document}